\documentclass[aps,prd,showpacs,showkeys,preprintnumbers,superscriptaddress,nobibnotes,floatfix,longbibliography,nofootinbib]{revtex4-1}
\pdfoutput=1
\usepackage{amsmath}
\usepackage{amsfonts}
\usepackage{amssymb}
\usepackage{mathrsfs}
\usepackage{graphicx}
\usepackage{color}
\usepackage{longtable}
\usepackage{hyperref}
\usepackage{multirow}
\usepackage{slashed}
\usepackage{epsf}

\usepackage{color}

\begin{document}

\title{
Probing the Supersymmetric Grand Unified Theories with Gravity Mediation 
at the Future Proton-Proton Colliders and Hyper-Kamiokande Experiment
}

\author{Waqas Ahmed}
\email{waqasmit@hbpu.edu.cn}
\affiliation{School of Mathematics and Physics, Hubei Polytechnic University, Huangshi 435003,China }
\author{Tianjun Li}
\email{tli@mail.itp.ac.cn}
\affiliation{ CAS Key Laboratory of Theoretical Physics, Institute of Theoretical Physics, Chinese Academy of Sciences, Beijing 100190, China}
\affiliation{School of Physical Sciences, University of Chinese Academy of Sciences, No. 19A Yuquan Road, Beijing 100049, China}
\author{Shabbar Raza}
\email{shabbar.raza@fuuast.edu.pk}
\affiliation{Department of Physics, Federal Urdu University of Arts, Science and Technology, Karachi 75300, Pakistan}


\begin{abstract}

With the grand desert hypothesis, we have proposed to probe the supersymmetric Grand Unified Theories (GUTs) 
at the future proton-proton (pp) colliders and Hyper-Kamiokande experiment previously. 
In this paper, we study the supersymmetric GUTs with gravity mediated supersymmetry breaking
in details. First, considering the dimension-six proton decay via heavy gauge boson exchange,
we point out that we can probe the supersymmetric GUTs with
 GUT scale $M_{GUT}$ up to $1.778\times 10^{16}$~GeV at the Hyper-Kamiokande experiment.
Second, for the supersymmetric GUTs with $M_{GUT} \ge 1.0\times 10^{16}$~GeV and 
$M_{GUT} \ge 1.2\times 10^{16}$~GeV, we show that the upper bounds on the universal gaugino mass
are  $7.2$ TeV and 3.5 TeV, respectively, and thus the corresponding upper bounds on gluino mass
are  15 TeV and 8 TeV,  respectively. Also, we shall study the masses for charginos, neutralinos, 
squarks, sleptons, and Higgs particles in details. In particular, the supersymmetric GUTs with
$M_{GUT} \leq 1.2\times 10^{16}$~GeV can be probed at the Hyper-Kamiokande experiment,
and   the supersymmetric GUTs with $M_{GUT}\ge  1.2\times 10^{16}$~GeV can be probed at 
the future 100 TeV pp collider experiments such as the ${\rm FCC}_{\rm hh}$ and SppC via gluino searches. 
Thus, the  supersymmetric GUTs with gravity mediation can be probed by the ${\rm FCC}_{\rm hh}$, SppC,
and Hyper-Kamiokande experiments. In our previous study, we have shown that
the supersymmetric GUTs with anomaly and gauge mediated supersymmetry breakings are well within the reaches of 
these experiments. Therefore, our proposal provides the concrete scientific goal for 
the ${\rm FCC}_{\rm hh}$, SppC, and Hyper-Kamiokande experiments: probing the supersymmetric GUTs.

\end{abstract}
\maketitle

\section{Introduction}

It is well known that supersymmetry (SUSY) provides a natural solution to the gauge hierarchy problem 
in the Standard Model (SM). In the supersymmetric SMs (SSMs) with $Z_2$ $R$-parity,
we can achieve the gauge coupling unification~\cite{gaugeunification}, have the Lightest Supersymmetric Particle (LSP) like
neutralino as a dark matter (DM) candidate~\cite{Jungman:1995df}, and break the electroweak (EW) gauge symmetry radiatively 
due to the large top quark Yukawa coupling, etc. In particular,
gauge coupling unification strongly suggests 
the Grand Unified Theories (GUTs)~\cite{Georgi:1974sy,Pati:1974yy,Fritzsch:1974nn,Georgi:1974my}, 
and the SUSY GUTs can be constructed from  superstring theory, which is the
most competitive candidate for quantum gravity. 
Therefore, supersymmetry is not only the most promising new physics beyond the SM,
but also a bridge between the low energy phenomenology and high-energy
fundamental physics.

However, after accumulation of data from the LHC Run-1 and Run-2, we have no hints for the SSMs. Of course, 
with the help of theses data now we have stronger bounds on the spectra of the supersymmetric particles (sparticles). 
For instance, the lower mass bounds on gluino, 
the first two generation of squarks, stop, and sbottom are  2.3 TeV, 1.9 TeV, 1.25 TeV, and 1.5 TeV, 
respectively~\cite{ATLAS-SUSY-Search, Aad:2020sgw, Aad:2019pfy, CMS-SUSY-Search-I, CMS-SUSY-Search-II}. 
Thus, there might exist the EW fine-tuning problem, and some promising solutions have been 
proposed~\cite{Dimopoulos:1995mi, Cohen:1996vb, Kitano:2006gv, LeCompte:2011cn, Fan:2011yu, Kribs:2012gx, 
Baer:2012mv, Ding:2015epa,Batell:2015fma,Leggett:2014hha,Du:2015una,Li:2015dil,Ellis:1986yg,Barbieri:1987fn}. 
These natural SSMs generically predicts some relatively light sparticles, for
instance, Bino, Higgsino, stop, gluino, and sleptons, etc.

On the other hand, to probe the new physics beyond the SM, we have 
a few proposals for the future proton-proton (pp) colliders, for example, 
the ${\rm FCC}_{\rm hh}$~\cite{Benedikt:2018csr}  and SppC~\cite{CEPC-SPPCStudyGroup:2015csa}. 
The naive question is whether
we can probe the supersymmetry at the ${\rm FCC}_{\rm hh}$ and SppC, but
the answer is no since the sparticles can be very heavy.
And then the interesting question is whether we can probe the supersymmetric GUTs at the future experiments
even if there does exist the SUSY EWFT problem.
In a previous study~\cite{Ahmed:2020qdt}, with the grand desert hypothesis, we showed that 
the supersymmetric GUTs can be probed at the future proton-proton (pp) colliders and 
Hyper-Kamiokande experiment. For the GUTs with the GUT scale $M_{GUT} \le 1.0\times 10^{16}$~GeV,
 the dimension-six proton decay via heavy gauge boson exchange can be probed
at the Hyper-Kamiokande experiment. Also, for the GUTs with 
$M_{GUT} \ge 1.0\times 10^{16}$~GeV, we showed that the GUTs with anomaly and 
gauge mediated supersymmetry breakings are well within the reaches of 
the future 100 TeV pp colliders such as the ${\rm FCC}_{\rm hh}$ and SppC,
and the supersymmetric GUTs with gravity mediated supersymmetry breaking 
can be probed at the future 160 TeV pp collider. Therefore, the
remaining interesting question is whether we can probe the
supersymmetric GUTs with gravity mediated supersymmetry breaking at the 
${\rm FCC}_{\rm hh}$ and SppC.

In this paper, we shall study the supersymmetric GUTs with gravity mediated supersymmetry breaking~\cite{chams, bbo, cmssm} 
in details. 
First, considering the dimension-six proton decay via heavy gauge boson exchange,
we point out that we can probe the supersymmetric GUTs with
 GUT scale $M_{GUT}$ up to $1.778\times 10^{16}$~GeV at the Hyper-Kamiokande experiment.
Second, for the supersymmetric GUTs with $M_{GUT} \ge 1.0\times 10^{16}$~GeV and 
$M_{GUT} \ge 1.2\times 10^{16}$~GeV, we show that the upper bounds on the universal gaugino mass
are  $7.2$ TeV and 3.5 TeV, respectively, and thus the corresponding upper bounds on gluino mass
are  15 TeV and 8 TeV,  respectively. Also, we shall study the masses for charginos, neutralinos, 
squarks, sleptons, and Higgs particles in details. In particular, the supersymmetric GUTs with
$M_{GUT} \leq 1.2\times 10^{16}$~GeV can be probed at the Hyper-Kamiokande experiment,
and   the supersymmetric GUTs with $M_{GUT}\ge  1.2\times 10^{16}$~GeV can be probed at 
the  ${\rm FCC}_{\rm hh}$ and SppC experiments via gluino searches. 
Thus, the  supersymmetric GUTs with gravity mediation can be probed by the ${\rm FCC}_{\rm hh}$, SppC,
and Hyper-Kamiokande experiments. In our previous study, we have shown that
the supersymmetric GUTs with anomaly and gauge mediated supersymmetry breakings are well within the reaches of 
these experiments~\cite{Ahmed:2020qdt}. Therefore, we propose the concrete scientific goal for 
the ${\rm FCC}_{\rm hh}$, SppC, and Hyper-Kamiokande experiments: probing the supersymmetric GUTs.

This paper is organized as follows. In Section II, we discuss the
supersymmetric GUT searches at the Hyper-Kamiokande experiment.
In Section III, we study the supersymmetric GUTs  
with gravity mediated supersymmetry breaking in details,
and their searches at the future proton-proton colliders.
Our conclusion is given in Section IV.

\section{Probing the Supersymmetric Grand Unified Theories at the Hyper-Kamiokande Experiment }

In the GUTs, the well-know prediction is the dimension-six proton decay $p\to e^+ \pi^0$ via heavy 
gauge boson exchange, and the proton lifetime is given by~\cite{Hisano:2000dg,Dutta:2016jqn}
\begin{eqnarray}
\tau_p(e^+\pi^0) &\simeq& 1.0\times 10^{34} \times
\left(\frac{2.5}{A_R}\right)^2 \times
\left(\frac{0.04}{\alpha_{\rm GUT}}\right)^2
\nonumber \\  &&
\times
\left(\frac{M_{\rm X/Y}}{1.0\times 10^{16}~{\rm GeV}}\right)^4
~{\rm years} ~,~\,
\label{eq:proton}
\end{eqnarray}
where $A_R$ is  the dimensionless one-loop renormalization factor associated with anomalous dimension 
of the relevant baryon-number violating operators, $\alpha_{\rm GUT}$ is the unified gauge coupling,
and $M_{\rm X/Y}$ is the mass for the heavy gauge bosons $X_{\mu}/Y_{\mu}$. 
The current lower limit on the proton lifetime from
the Super-Kamiokande experiment is $\tau_p > 1.6 \times 10^{34}$ years~\cite{Miura:2016krn}.
Thus, we obtain $M_{\rm X/Y} \ge 1.0\times 10^{16}~{\rm GeV}$.
At the future Hyper-Kamiokande experiment, we can probe the proton lifetime
at least above $1.0 \times 10^{35}$ years~\cite{Abe:2018uyc}.
Thus, the GUTs with $M_{\rm X/Y} \le 1.778\times 10^{16}~{\rm GeV}$ is within the reach of 
the future Hyper-Kamiokande experiment. For more detail related GUTs and $M_{GUT}$ related bosons see~\cite{Hisano:2013cqa, Ellis:2015jwa, Pokorski:2019ete,Tobe:2003yj,Goto:1998qg, Murayama:2001ur,Liu:2013ula,Hisano:2013exa,Babu:2020ncc}.

To clarify the subtle point, we want to emphasize that the mass of the heavy gauge bosons $X_{\mu}/Y_{\mu}$
is smaller than or equal to the GUT scale  $M_{GUT}$. Thus, the supersymmetric GUTs with GUT scale 
up to $1.778\times 10^{16}~{\rm GeV}$ can be probed at the future Hyper-Kamiokande experiment.


\section{Probing the Supersymmetric Grand Unified Theories with Gravity Mediation 
at the Future Proton-Proton Colliders }\label{sec:scan}

The supersymmetry searches at the 100~TeV pp colliders have been studied previously~\cite{Benedikt:2018csr,CEPC-SPPCStudyGroup:2015csa, Cohen:2013xda, Arkani-Hamed:2015vfh,
Fan:2017rse, Golling:2016gvc}. For the integrated luminosity 30~${\rm ab}^{-1}$,
Wino via Bino decay, gluino ${\tilde g}$ via heavy flavor decay, gluino via light flavor decay,
first-two generation squarks ${\tilde q}$, and stop can be discovered for their masses
up to about 6.5~TeV, 11~TeV, 17~TeV, 14~TeV, and 11~TeV, respectively.
Moreover, if the gluino and first-two generation squark masses are similar,
they can be probed up to 20 TeV.

 \begin{figure}[t!]
\includegraphics[width=1.0\columnwidth]{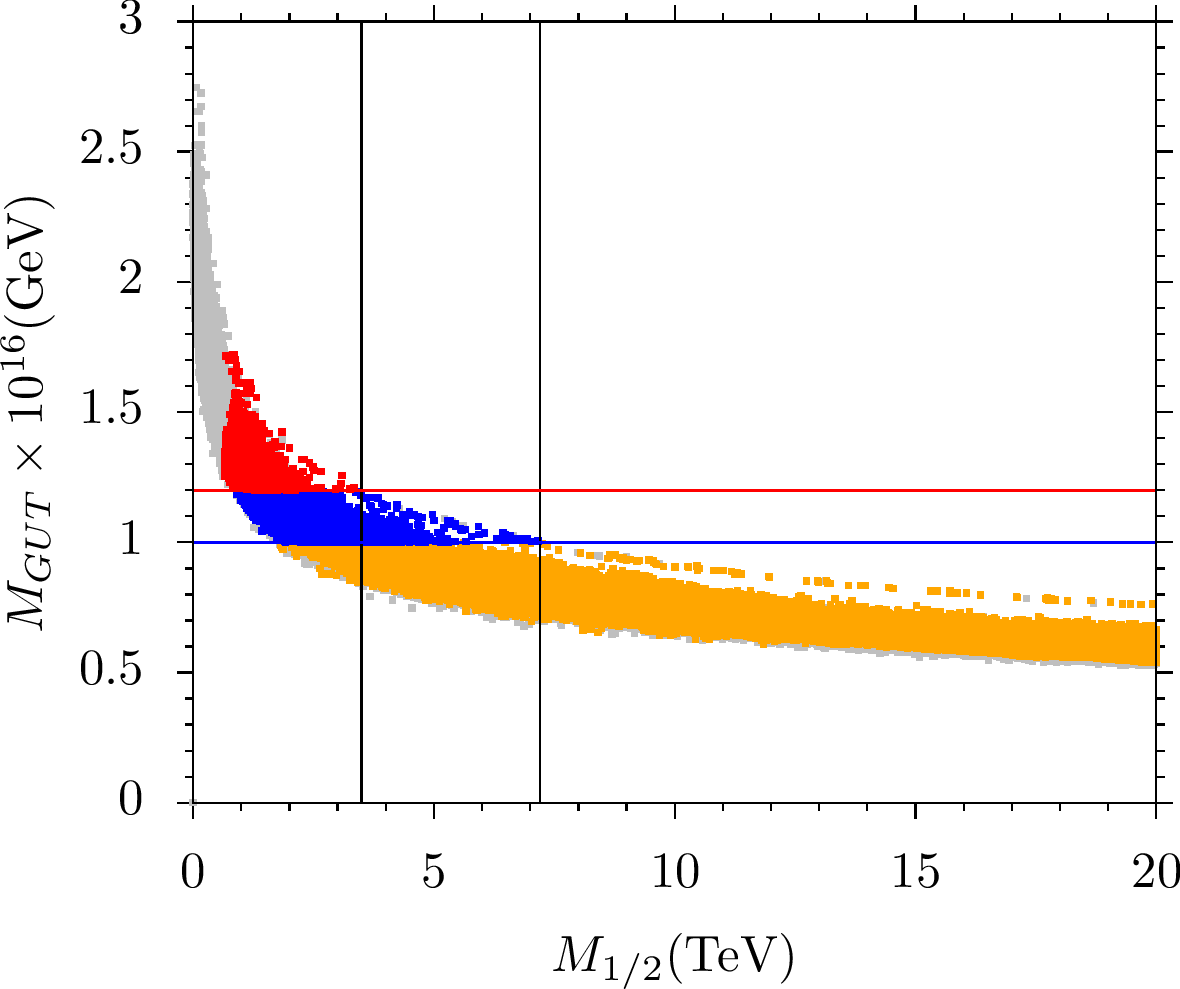}
\caption{Plot in $M_{1/2}-M_{GUT}$ plane. Gray points are consistent with REWSB and LSP neutralino. 
Orange points satisfy the mass bounds including $m_{h}=125 \pm 3\,{\rm GeV}$
and the constraints from rare $B-$ meson decays. Blue points form a subset of orange points and satisfy 
$1 \lesssim M_{GUT}\lesssim 1\times 10^{16} \,{\rm GeV}$, while red points form a subset of orange points 
and satisfy $M_{GUT} \gtrsim 1.2\times 10^{16}\, {\rm GeV}$. Two horizontal blue and red lines represent 
$M_{GUT}=1\times 10^{16}\,{\rm GeV}$ and $M_{GUT}=1.2\times 10^{16}\, {\rm GeV}$, respectively. 
The first vertical  line shows the upper bound on $M_{1/2}$ for red points ($M_{1/2}= 3.5$ TeV), and the second vertical line 
shows the upper bound on $M_{1/2}$ for blue points ($M_{1/2}=7.2$ TeV). }
\label{fig1}
\end{figure}

By the way, the correlations between the low energy SUSY spectra
and the GUT scale have been studied via the one-loop renormalization group
equations before, and it was found that the bound from dimension-six proton decay
already excludes the gluinos and Winos heavier than about 120 TeV and 40 TeV, respectively,
if their mass ratio $M_3/M_2$ is about 3~\cite{Pokorski:2017ueo}. In our paper, we employ 
the  ISAJET~7.85 package~\cite{ISAJET} to 
perform the scan, which will give us more precise results. To be concrete,
for the supersymmetric GUTs with $M_{GUT} \le 1.0\times 10^{16}$~GeV,
 we find that the current bound from dimension-six proton decay
 excludes the gluinos and Winos heavier than 15 TeV and 6 TeV, respectively.
Because  Winos might decay via Higgsinos as the benchmark point 4 given in the following subsection B,
we are not sure whether Wino is within the reach of the  ${\rm FCC}_{\rm hh}$ and SppC experiments,
which will be studied elsewhere. Moreover,
 we shall study the masses for charginos, neutralinos, squarks, sleptons, and Higgs particles as well.

\subsection{Phenomenological Constraints and Scanning Procedure}

 We summarize our scanning procedure and the experimental constraints in this part of our article. The package ISAJET~7.85 ~\cite{ISAJET} is employed to perform the random scans over the parameter space of mSUGRA/CMSSM. In this code, the weak scale values of the gauge and third
 generation Yukawa couplings are evolved to
 $M_{\rm GUT}$ via the MSSM RGEs
 in the $\overline{DR}$ regularization scheme.
It is to be noted that we do not strictly enforce the unification condition
 $g_3=g_1=g_2$ at $M_{\rm GUT}$. It is because a few percent deviation
 from unification can be assigned to the unknown GUT-scale threshold
 corrections~\cite{Hisano:1992jj}. {Moreover we do not allow $g_{3}$ to deviate from the unification by more than about $3\%$.}
We define the boundary conditions of our model at $M_{\rm GUT}$ and then  all the Soft Supersymmetry Breaking (SSB) parameters, 
along with the gauge and Yukawa couplings,  are evolved back to the weak scale $M_{\rm Z}$.

In evaluating Yukawa couplings, the SUSY threshold
 corrections~\cite{Pierce:1996zz} are taken into account
 at the common scale $M_{\rm SUSY}= \sqrt{m_{\tilde t_{L}}m_{\tilde t_{R}}}$.
The entire parameter set is iteratively run between
 $M_{\rm Z}$ and $M_{\rm GUT}$ using the full two-loop RGEs
 until a stable solution is obtained.
To better account for the leading-log corrections, one-loop step-beta
 functions are adopted for the gauge and Yukawa couplings, and
 the SSB parameters $m_i$ are extracted from RGEs at appropriate scales {as}
 $m_i=m_i(m_i)$.
The RGE-improved one-loop effective potential is minimized
 at an optimized scale $M_{\rm SUSY}$, which effectively
 accounts for the leading two-loop corrections.
The full one-loop radiative corrections are incorporated
 for all sparticles.
 
 Note that we set $\mu > 0$ and  use $m_t = 173.3\, {\rm GeV}$  \cite{:2009ec}.
We also use $m_b^{\overline{DR}}(M_{\rm Z})=2.83$ GeV which is hard-coded into ISAJET.  
\begin{figure*}[ht]
    \centering
        \begin{tabular}{c c}
    \includegraphics[width = 0.5\textwidth]{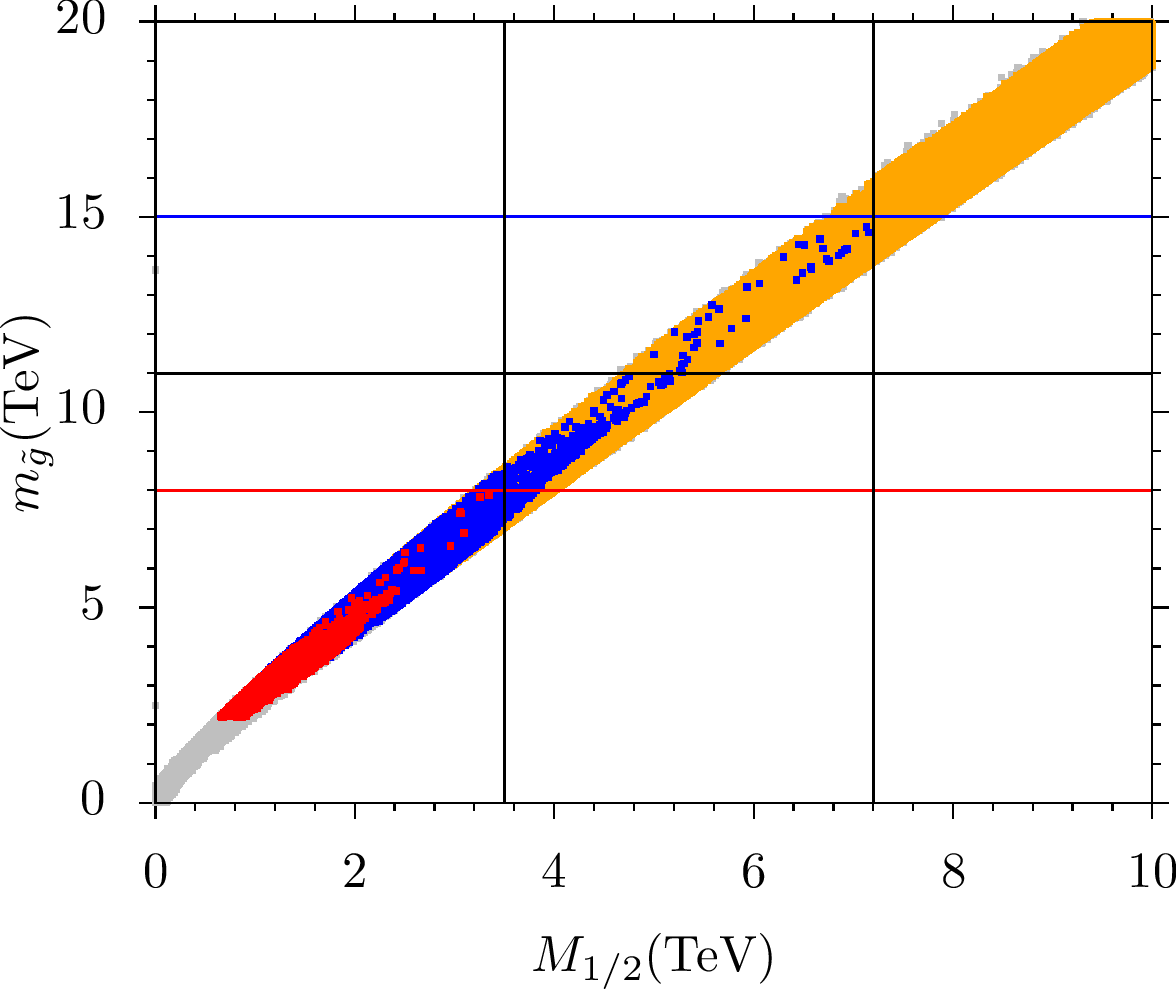}  
 \hspace{-.01cm}
\includegraphics[width = 0.5\textwidth]{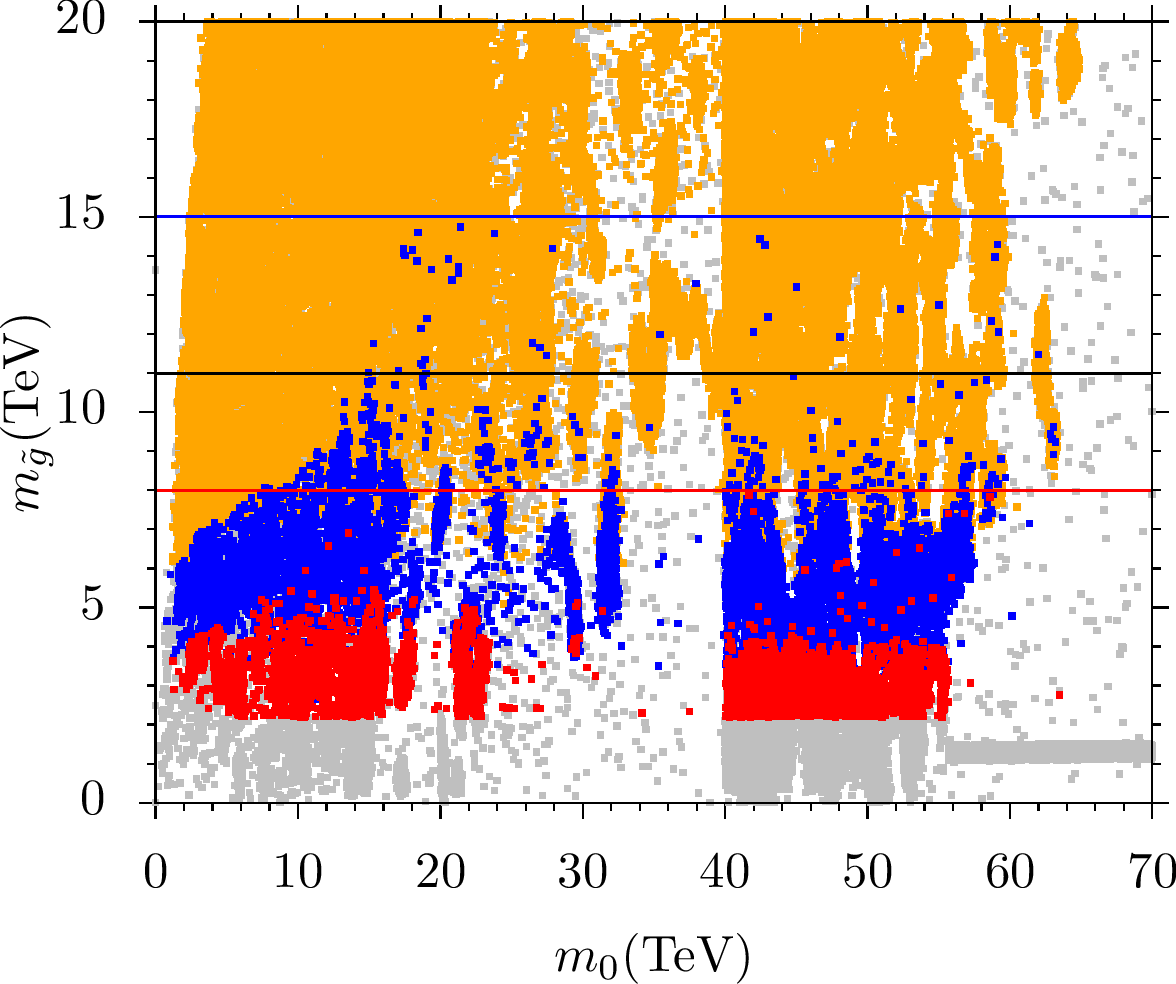} 

\\
\includegraphics[width = 0.5\textwidth]{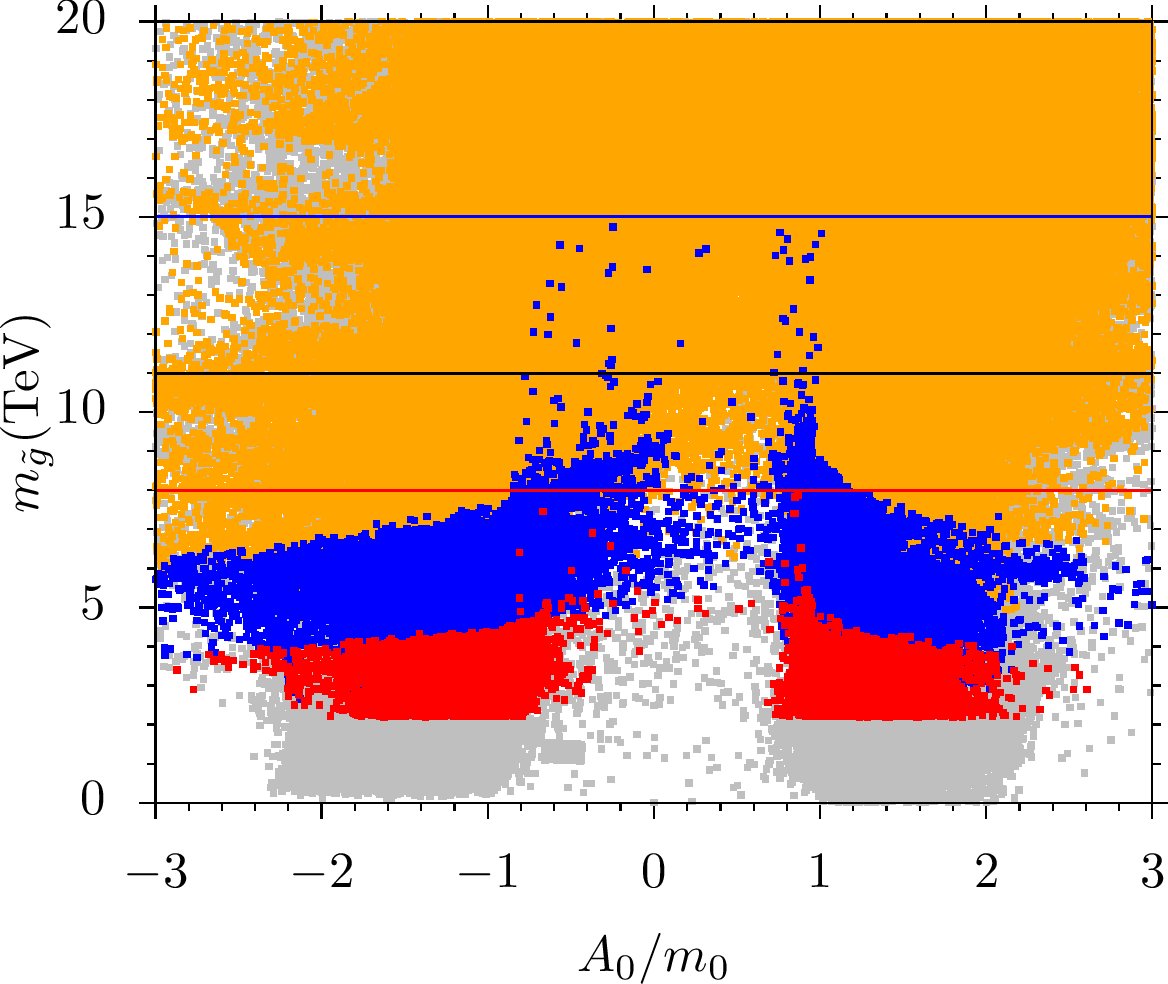}  
 \hspace{-.01cm}
\includegraphics[width = 0.5\textwidth]{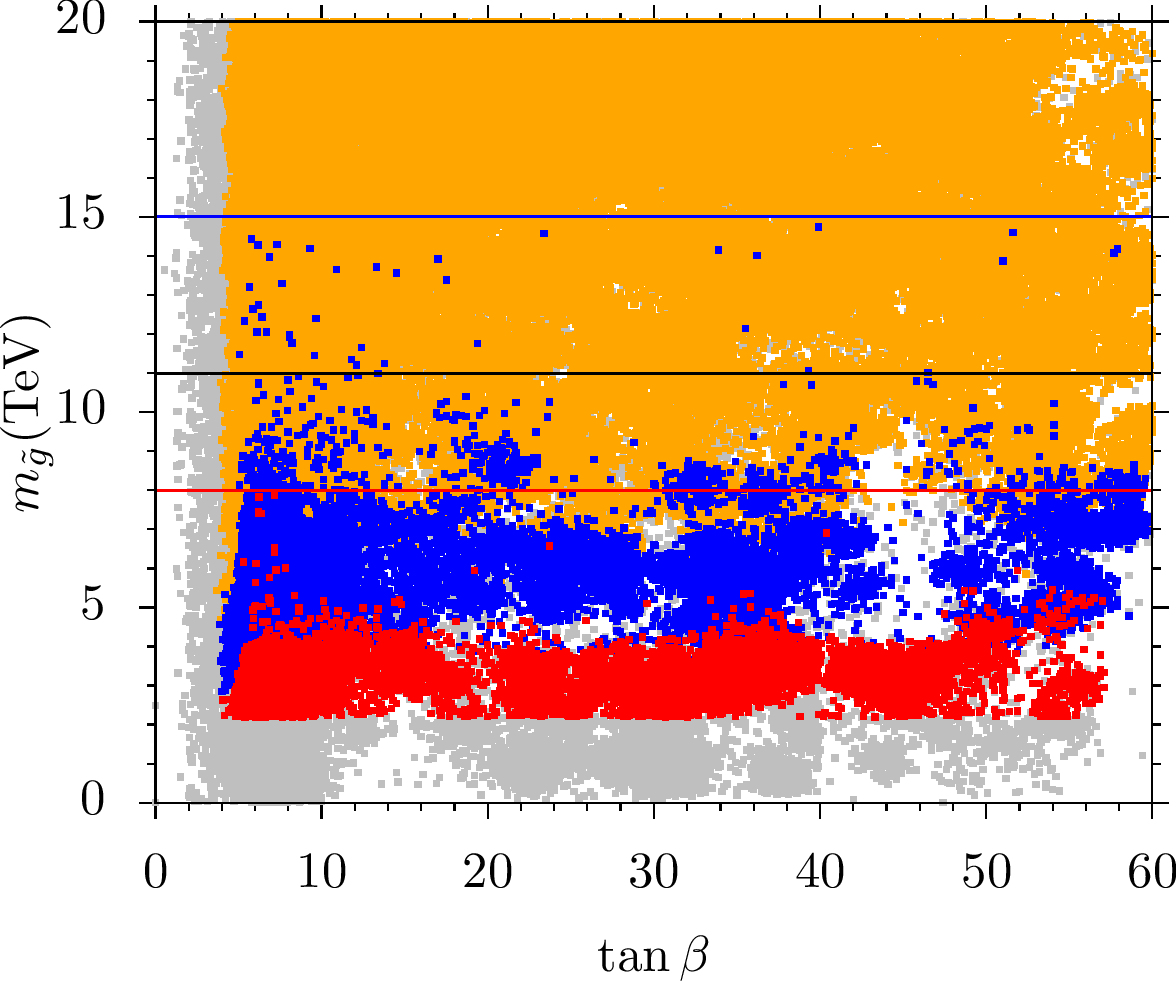} 
    \end{tabular}
    \caption{The gluino mass versus supersymmetry breaking soft terms and $\tan\beta$.
The color coding is the same as in Fig.~\ref{fig1}.
 Horizontal red, black and blue lines  represent 
the gluino mass upper bounds of 8 TeV, 11 TeV, and 15 TeV corresponding to the red points, 
gluino discovery via heavy flavor decay at 100 TeV pp collider~\cite{Cohen:2013xda}, and blue points, respectively. 
The first vertical  line shows the upper bound on $M_{1/2}$ for red points ($M_{1/2}= 3.5$ TeV), and the second vertical line 
shows the upper bound on $M_{1/2}$ for blue points ($M_{1/2}=7.2$ TeV). }
    \label{fig2}
\end{figure*}
The fundamental parameters of mSUGRA/CMSSM are restricted as follows

\begin{align}
0\leq & m_{0}  \leq 90 ~\rm{TeV}   ~,~\nonumber \\
0\leq & M_{1/2}  \leq 30 ~\rm{TeV}  ~,~\nonumber \\
-3\leq & A_{0}/m_{0}  \leq 3   ~,~\nonumber \\
2\leq & \tan\beta  \leq 60   ~.~\nonumber \\
 \label{input_param_range}
\end{align}

We would like to draw attention of the reader to the fact that the requirement of radiative electroweak symmetry breaking (REWSB)~\cite{Ibanez:1982fr} puts an important theoretical
 constraint on the parameter space.
Another important constraint comes from the limits on the cosmological
 abundance of stable charged {particles}~\cite{Beringer:1900zz}.
They exclude the parameter space where the charged
 SUSY particles, such as $\tilde \tau_{1}$ or $\tilde t_{1}$,
 become the LSP.

The data points collected all satisfy the requirement of REWSB, with the neutralino being the LSP. In addition, after collecting the data, we impose the mass bounds on all the sparticles \cite{Agashe:2014kda}, and the constraints from rare decay processes $B_{s}\rightarrow \mu^{+}\mu^{-} $ \cite{Aaij:2012nna}, $b\rightarrow s \gamma$ \cite{Amhis:2012bh}, and $B_{u}\rightarrow \tau\nu_{\tau}$ \cite{Asner:2010qj}. 
More explicitly, we set
\begin{eqnarray}
m_h  = 122-128~{\rm GeV}~~&,
\\
m_{\tilde{g}}\geq 2.2~{\rm TeV}~, \\
0.8\times 10^{-9} \leq{\rm BR}(B_s \rightarrow \mu^+ \mu^-)
  \leq 6.2 \times10^{-9} \;(2\sigma)~~&&,
\\
2.99 \times 10^{-4} \leq
  {\rm BR}(b \rightarrow s \gamma)
  \leq 3.87 \times 10^{-4} \; (2\sigma)~~&&,
\\
0.15 \leq \frac{
 {\rm BR}(B_u\rightarrow\tau \nu_{\tau})_{\rm MSSM}}
 {{\rm BR}(B_u\rightarrow \tau \nu_{\tau})_{\rm SM}}
        \leq 2.41 \; (3\sigma)~~&&.
\end{eqnarray}
To be general, we do not require the relic abundance of the LSP neutralino to satisfy the Planck 2018 bound within 
$5\sigma$: $0.114 \leq \Omega_{\rm CDM}h^2 (\rm Planck) \leq 0.126$~\cite{Akrami:2018vks}.

\begin{figure*}[ht]
    \centering
        \begin{tabular}{c c}
    \includegraphics[width = 0.5\textwidth]{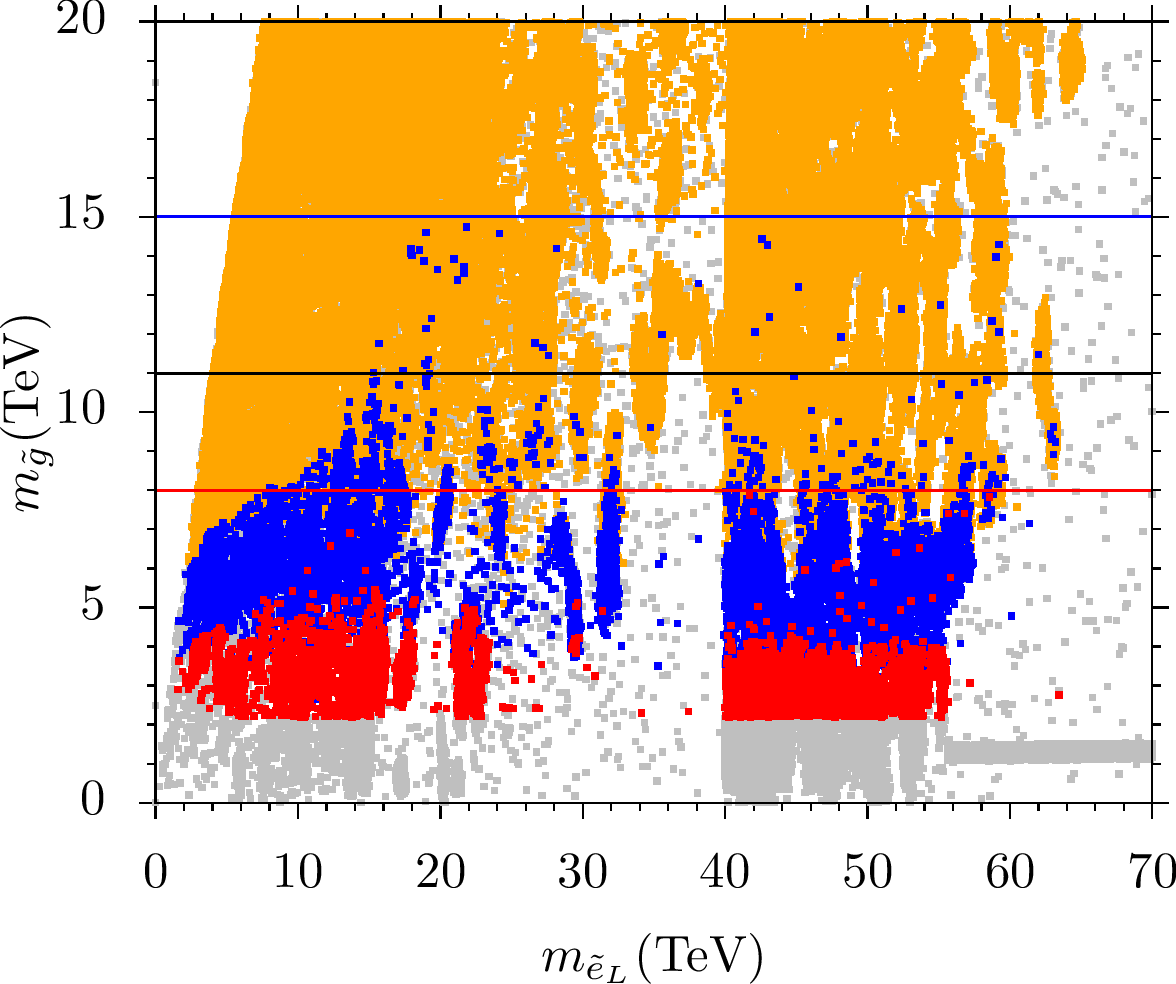}  
 \hspace{-.01cm}
\includegraphics[width = 0.5\textwidth]{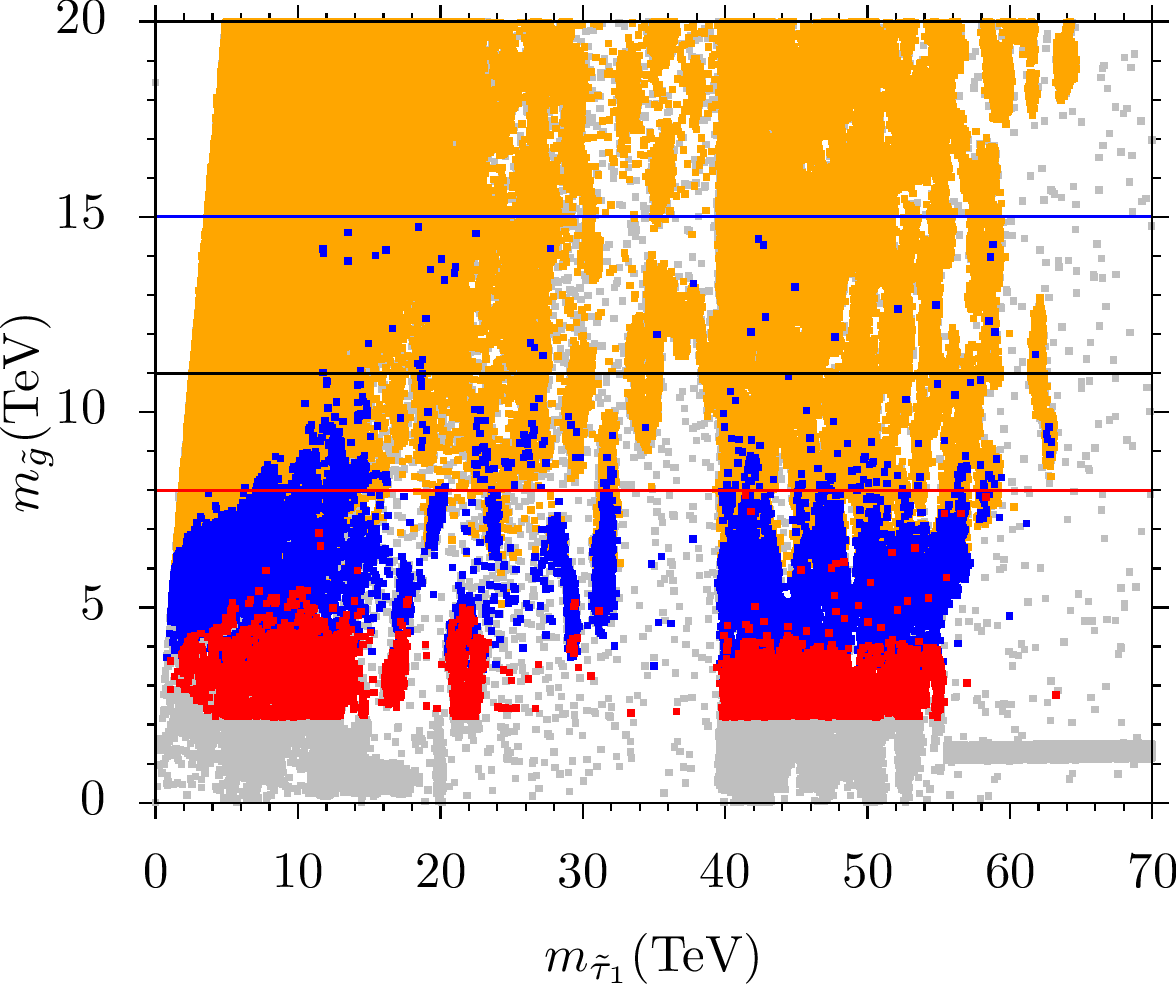} 

\\
   
\includegraphics[width = 0.5\textwidth]{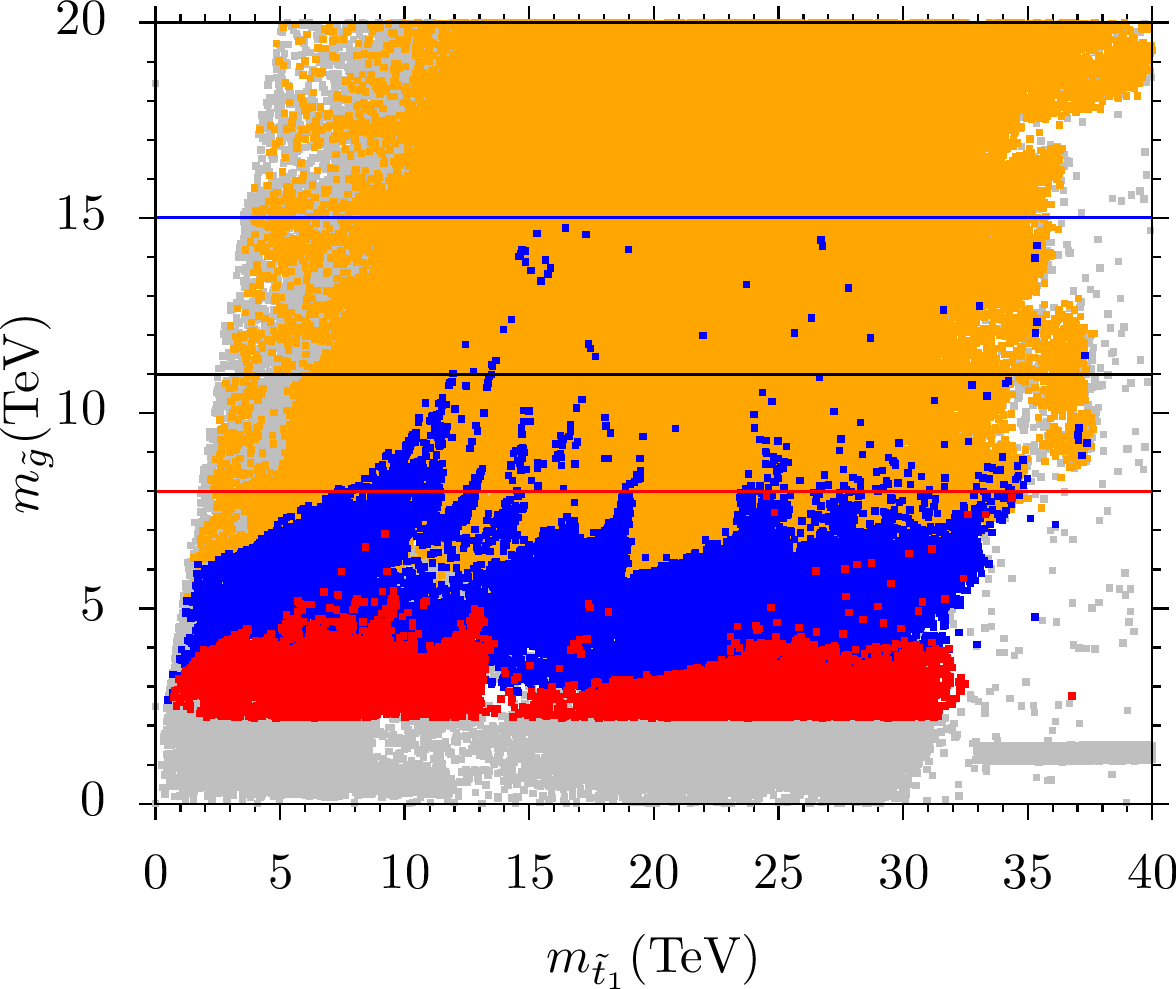}  
 \hspace{-.01cm}
\includegraphics[width = 0.5\textwidth]{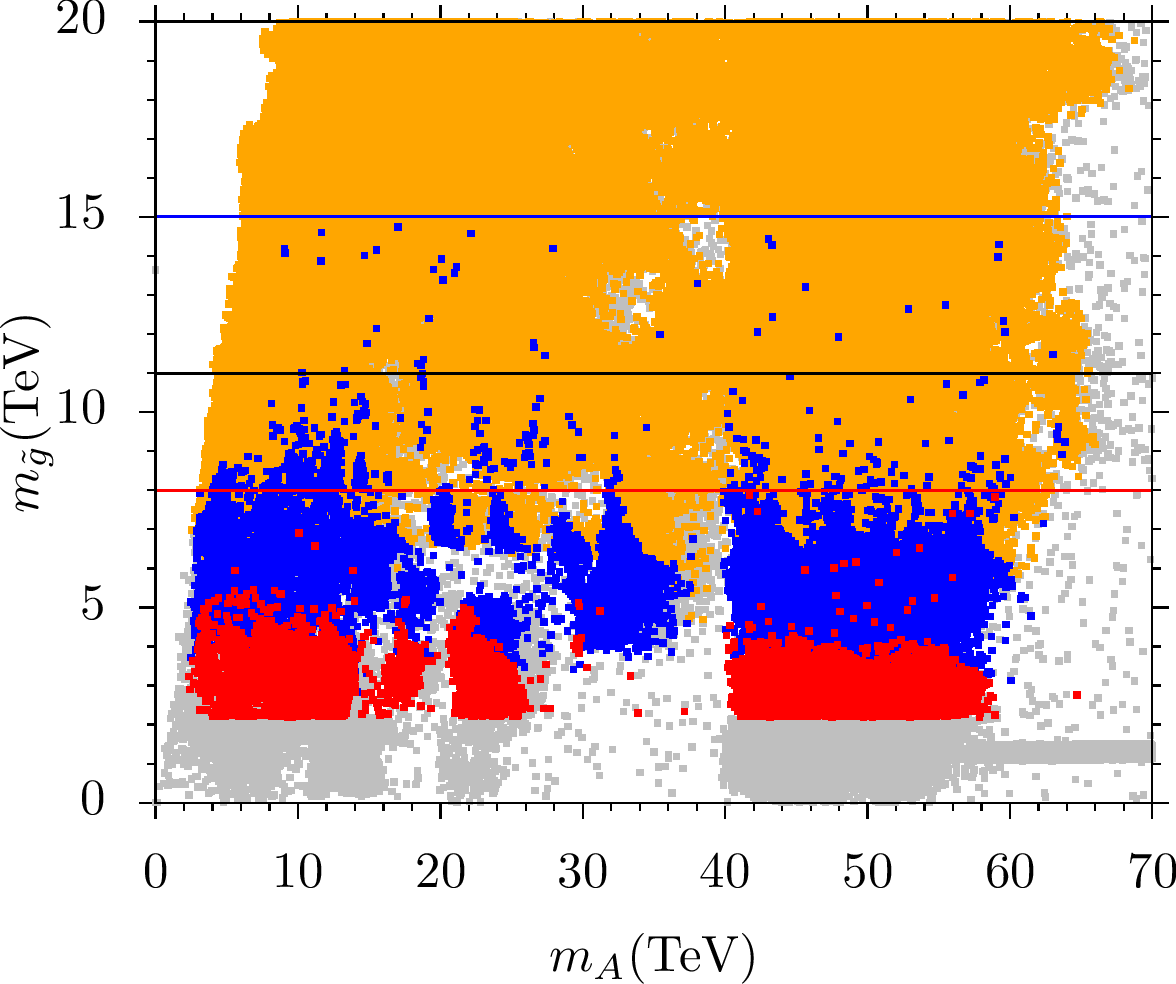}    
   
    \end{tabular}
    \caption{The gluino mass versus the left-handed slepton mass, light stau mass, light stop mass, 
and CP-odd Higgs boson mass. The color coding is the same as in Fig.~\ref{fig2}. 
}
    \label{fig3}
\end{figure*}

\begin{figure*}[ht]
    \centering
        \begin{tabular}{c c}
    \includegraphics[width = 0.5\textwidth]{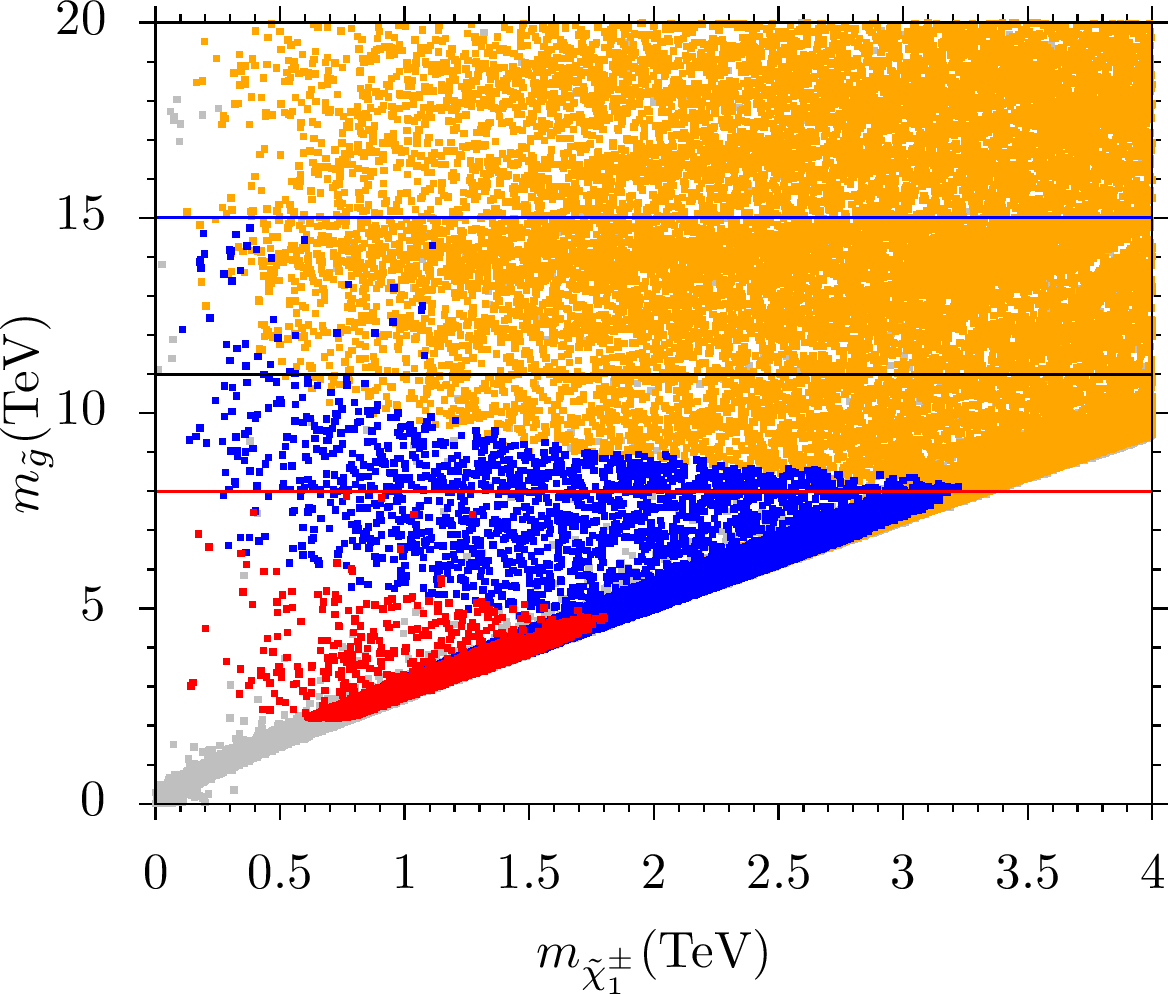}  
 \hspace{-.01cm}
\includegraphics[width = 0.5\textwidth]{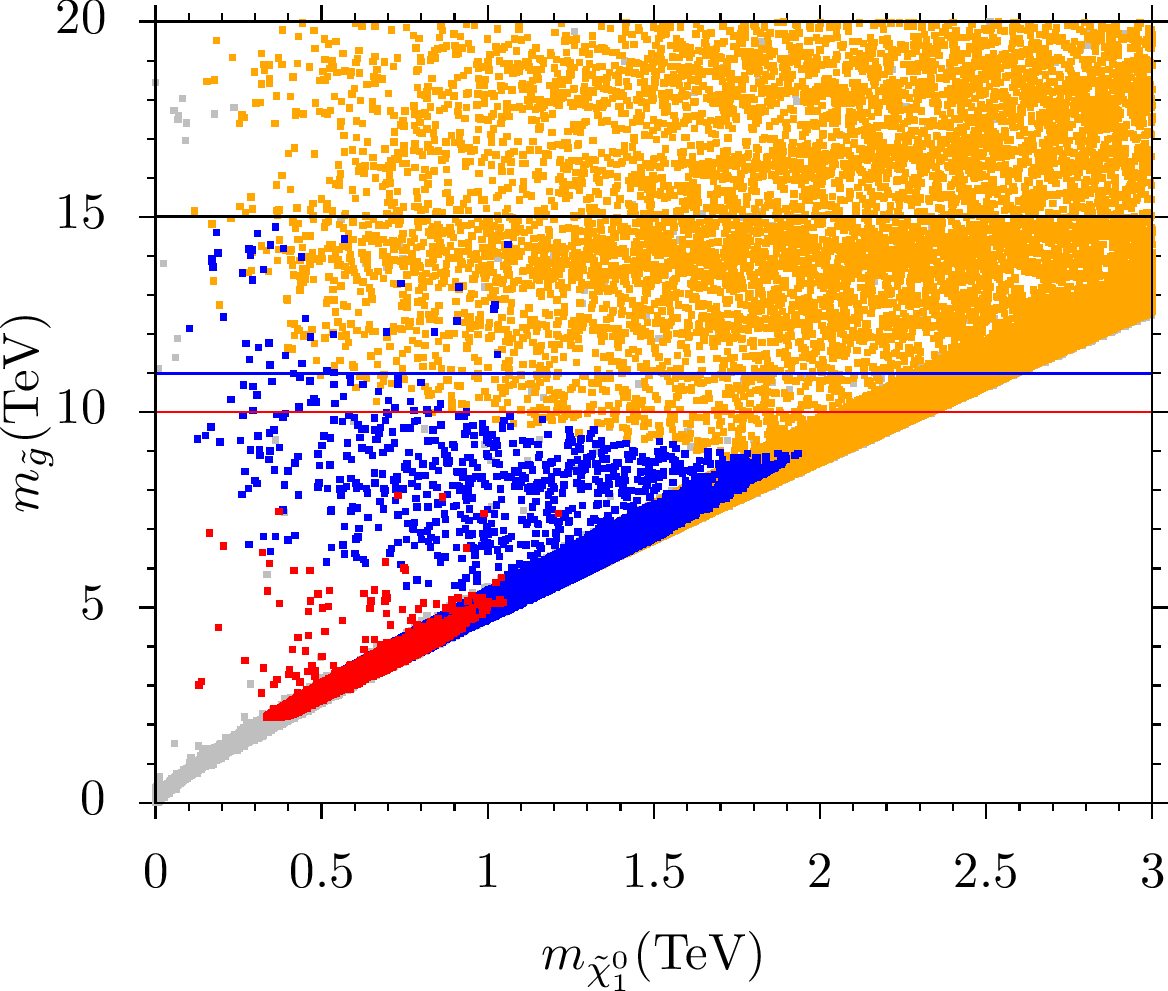} 

\\
   
\includegraphics[width = 0.5\textwidth]{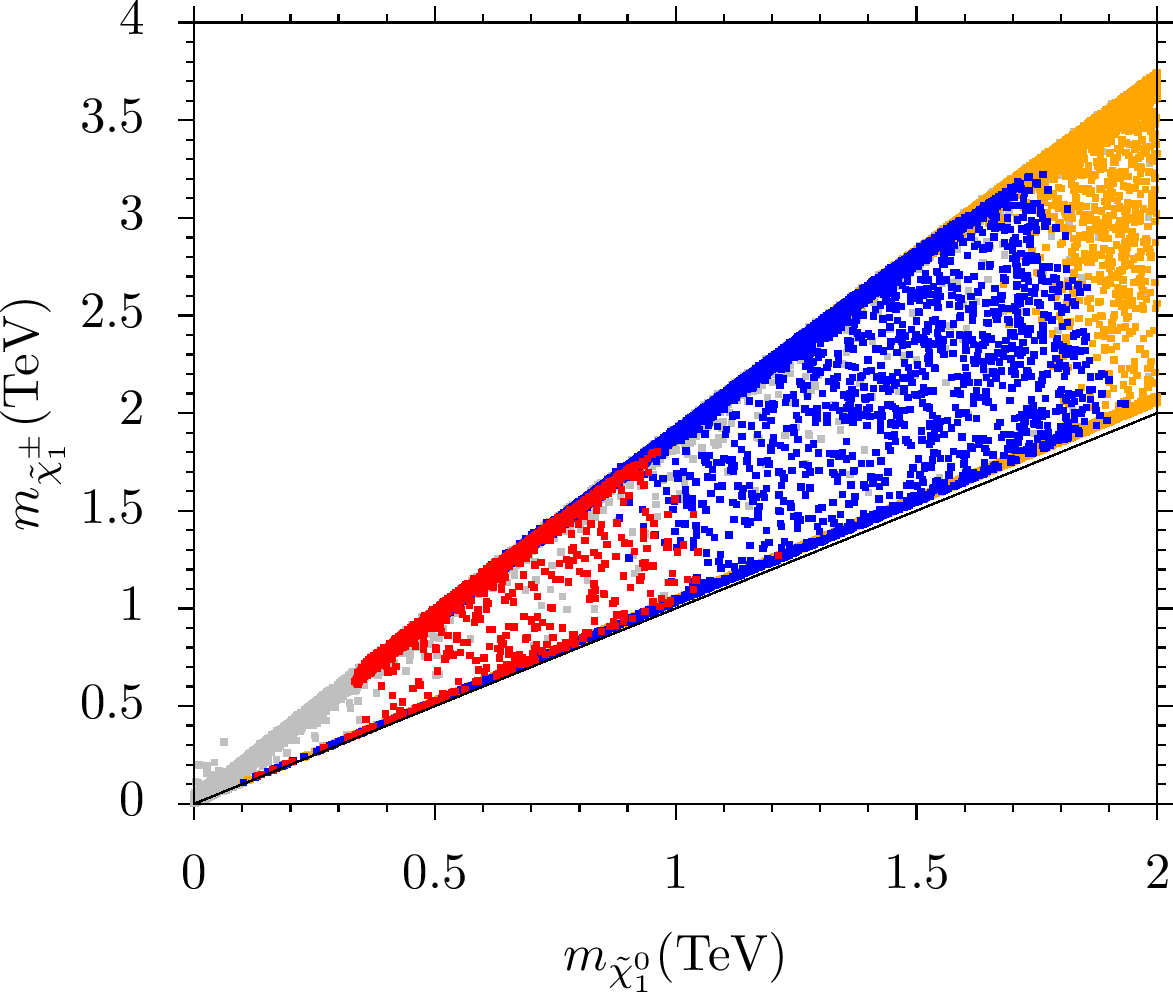}  
 \hspace{-.01cm}
\includegraphics[width = 0.5\textwidth]{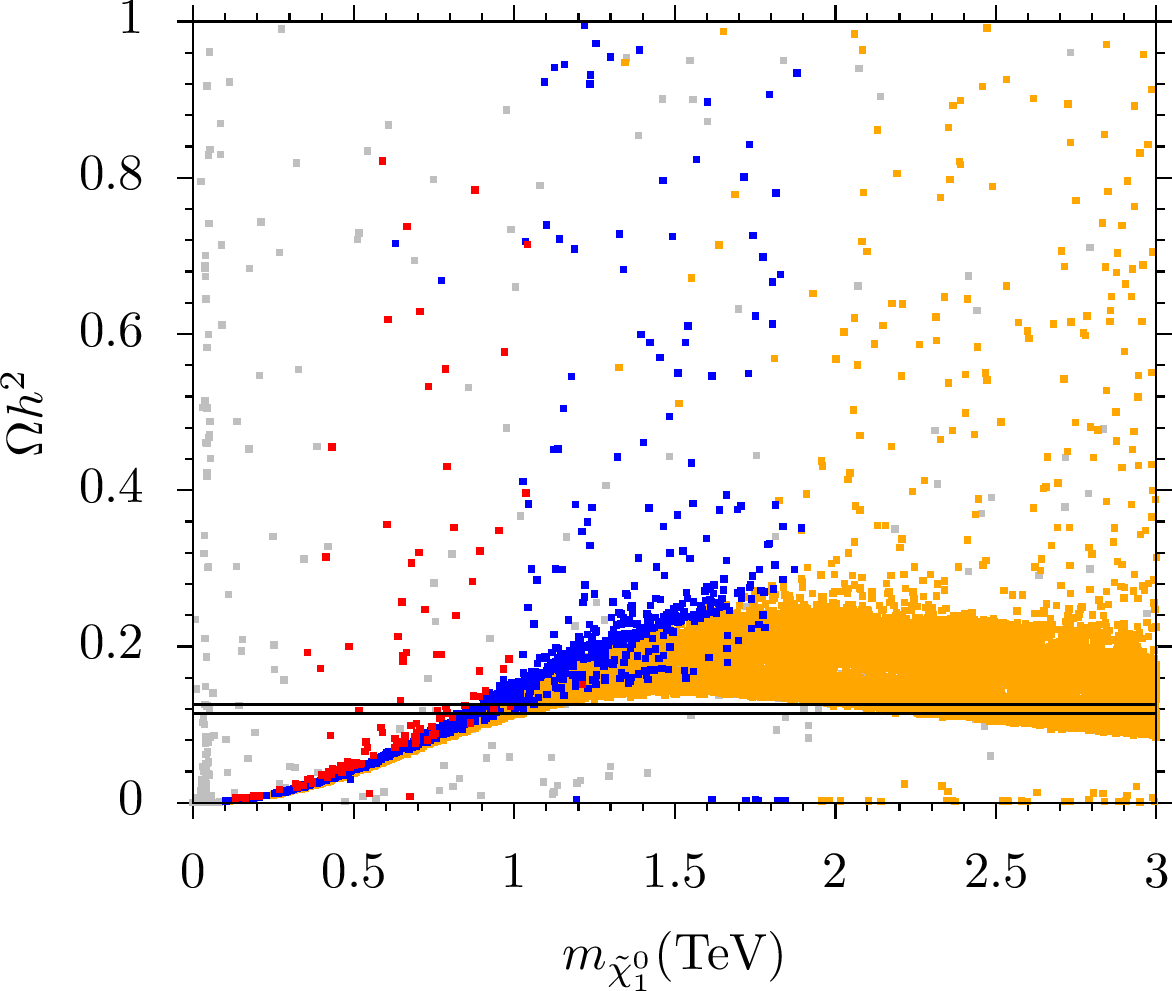}    
   
    \end{tabular}
    \caption{The gluino mass versus the light chargino mass and lightest neutralino mass,
the light chargino mass versus the lightest neutralino mass, and dark matter density
versus the lightest neutralino mass.
The color coding is the same as in Fig.~\ref{fig2}.}
    \label{fig4}
\end{figure*}


\subsection{Scan Results}\label{SR} 

We shall discuss results from the systematical scans. In Fig.~\ref{fig1},
 we show plot $M_{GUT}$ as a function of $M_{1/2}$. Gray points are consistent 
with REWSB and LSP neutralino. Orange points satisfy the mass bounds including
 $m_{h}=125 \pm 3\,{\rm GeV}$ and the constraints from rare $B-$ meson decays. 
Blue points form a subset of orange points and satisfy $1 \lesssim M_{U}\lesssim 1\times 10^{16} \,{\rm GeV}$,
 while red points form a subset of orange points and satisfy $M_{GUT} \gtrsim 1.2\times 10^{16}\, {\rm GeV}$. 
Two horizontal blue and red lines represent $M_{GUT}=1\times 10^{16}\,{\rm GeV}$ 
and $M_{GUT}=1.2\times 10^{16}\, {\rm GeV}$, respectively. 
The first vertical  line shows the upper bound on $M_{1/2}$ for red points ($M_{1/2}= 3.5$ TeV), while the second vertical line 
shows the upper bound on $M_{1/2}$ for blue points ($M_{1/2}=7.2$ TeV).
From the upper bounds on gaugino masses $M_{1/2}$ given by two vertical lines, we obtain that the upper bounds
on gluino masses are 8 TeV and 15 TeV respectively for the red and blue points. 
Therefore, we clearly show that SUSY GUTs with $M_{GUT} \gtrsim 1.2\times 10^{16}$~GeV 
for gravity mediated SUSY breaking scenario~\cite{chams, bbo, cmssm}, {\it i.e.} the red points, can be probed 
by the future 100 TeV pp colliders such as the ${\rm FCC}_{\rm hh}$ and SppC. 
Moreover, the blue points and orange points with $M_{GUT} \le 1.0\times 10^{16}\, {\rm GeV}$
 can be explored by the Hyper-Kamiokande experiment. 
In the latter part of the paper we see the impact of these bounds on the fundamental parameters 
of the mSUGRA/CMSSM and sparticle spectrum.

In Fig.~\ref{fig2}, we display plots the mSUGRA/CMSSM fundamental parameters as function of gluino mass. 
The color coding is the same as in Fig.~\ref{fig1}. 
 Horizontal red, black and blue lines  represent 
the gluino mass upper bounds of 8 TeV, 11 TeV, and 15 TeV corresponding to the red points, 
gluino discovery via heavy flavor decay at 100 TeV pp collider~\cite{Cohen:2013xda}, and blue points, respectively. 
The first vertical line shows the upper bound on $M_{1/2}$ for red points ($M_{1/2}= 3.5$ TeV), and the second vertical line 
shows the upper bound on $M_{1/2}$ for blue points ($M_{1/2}=7.2$ TeV). 
 In the top left corner the plot in $M_{1/2}-m_{\tilde g}$ plane is shown. We see that 
the gluino mass range for red points is from 2.3 TeV to 8 TeV. And the Hyper-Kamiokande experiment can probe gluino with masses 
in the range starting from 3 TeV to 15 TeV. In the right panel we show plot in $m_{0}-m_{\tilde g}$ plane. 
We see that $m_{0}$ can be as heavy as 64 TeV for red points as well as blue points. This has important implications. 
Because $m_{\tilde q}^{2}\approx m_{0}^{2}+ (6-7)M_{1/2}^{2}$ , $m_{\tilde e_{L}}^{2}\approx m_{0}^{2}+ 0.5 M_{1/2}^{2}$, and $m_{\tilde e_{R}}^{2}\approx m_{0}^{2}+ 0.15 M_{1/2}^{2}$ ~\cite{Baer:2006rs},  the large $m_{0}$ term will give the dominate contributions
to the squark and slepton masses. 
Plot in $A_{0}/m_{0}-m_{\tilde g}$ plane is shown in lower left panel. Here we see that for smaller values of $A_{0}/m_{0}$, gluino mass $m_{\tilde g}$ rises. Plot is almost symmetric along $A_{0}/m_{0}=0$,  and $m_{\tilde g}$ decreases as as $|A_{0}/m_{0}|$ increases. 
Plot in $\tan\beta-m_{\tilde g}$ plane is shown in the lower right corner. 
It is evident from the plot that the red points, blue points and orange points can have $\tan\beta$ from 2 to 60.

In Fig.~\ref{fig3}, we present the gluino mass versus the left-handed slepton mass, light stau mass, light stop mass, 
and CP-odd Higgs boson mass. In the top upper panel, we display plot in $m_{\tilde e_{L}}-m_{\tilde g}$ plane.  
As we stated earlier, large $m_{0}$ term can give dominant contributions to the squark and slepton masses,
 so this plot is very much similar to $m_{0}-m_{\tilde g}$. 
 We do not show the plot for $m_{\tilde e_{R}}-m_{\tilde g}$ because it is similar to $m_{\tilde e_{L}}-m_{\tilde g}$ plot. 
Similarly, we depict $m_{\tilde \tau_{1}}-m_{\tilde g}$ plot in the top right panel in  which also have similar feature to $m_{\tilde e_{L}}-m_{\tilde g}$. Plot in $m_{\tilde t_{1}}-m_{\tilde g}$ plane is shown in lower left panel. 
Here we see the similar trend but the mass ranges are reduced.  
Plot in lower right panel is shown in $m_{A}-m_{\tilde g}$ plane. 
Here we notice that $m_{A}$ can be as heavy as 64 TeV for both red and blue points.

In Fig.~\ref{fig4}, we show the gluino mass versus the light chargino mass and lightest neutralino mass,
the light chargino mass versus the lightest neutralino mass, and dark matter density
versus the lightest neutralino mass.
Plot in the top left panel shows a graph in $m_{\tilde \chi_{1}^{\pm}}-m_{\tilde g}$ plane. It shows that for red points 
the chargino mass $m_{\tilde \chi_{1}^{\pm}}$ can be in the range from 0.1 TeV to 2 TeV but for the blue points the chargino mass reaches 
up to 3.4 TeV. Ref.~\cite{ATL-PHYS-PUB-2021-007} reports the exclusion limits on $\tilde \chi_{1}^{+}\tilde \chi_{1}^{-}$ and $\tilde \chi_{1}^{\pm}\tilde \chi_{2}^{0}$ productions with $\tilde l$-mediated decays and the productions with the SM-boson-mediated decays,
 which require $m_{\tilde \chi_{1}^{\pm}}\gtrsim$ 700 GeV. This means that most of our points can satisfy these bounds. 
Moreover, we will show later that for some of the lighter solutions $m_{\tilde \chi_{1}^{\pm}}$ and $m_{\tilde \chi_{1}^{0}}$ are mass-degenerate and then can evade these constraints as well.  A plot in $m_{\tilde \chi_{1}^{0}}-m_{\tilde g}$ plane is shown in the top right panel. 
This plot also has similar features to the previous plot. Here, the neutralino mass can reached up to 1.1 TeV for red points and 2 TeV for blue points. In the lower left panel, we show plot in $m_{\tilde \chi_{1}^{0}}-m_{\tilde \chi_{1}^{\pm}}$ plane. The diagonal black line is just to show the region of chargino-neutralino coannihilation where  chargino and neutralino masses are degenerate, and most of these points
might have the Higgsino LSP. 
For red and blue points, the coannihilation region has the upper mass bounds on $m_{\tilde \chi_{1}^{0}}$ about 1 TeV and 1.9 TeV, 
respectively. The coannihilation region 
with lower mass range as stated before is safe from collider constraints. In the right lower panel, we displays the plot with the LSP neutralino mass versus dark matter relic density $\Omega h^{2}$. Horizontal two black lines represent Planck 2018 5$\sigma$ bounds on cold dark matter relic density as shown above. 
It can be seen that some of the red and blue points can satisfy the relic density bounds. In the following, 
we present four benchmark points in Table \ref{table1} for the red and blue points, which are consistent with
the dark matter relic density bounds.  

%
\begin{table}[h!]
\centering
\scalebox{0.8}{
\begin{tabular}{lcccc}
\hline
\hline
                 & Point 1 & Point 2 & Point 3 & Point 4       \\
\hline
$m_{0}$            &    7966     &  56830   &  13170   &   59140        \\
$M_{1/2}$          &   2225      &  3056    & 4223     & 6457 \\
$A_{0}/m_{0}$      &   0.5109    &  0.84    &-0.038    & 0.9733\\
$\tan\beta$        & 54.8        &   6.2    & 20.6     & 7.32\\
\hline
$M_{GUT}$         & 1.234$\times 10^{16}$ & 1.226$\times 10^{16}$   & 1.035$\times 10^{16}$    &   1.01 $\times 10^{16}$      \\
\hline
$m_h$            &  123    & 125    &  125   &   126     \\
$m_H$            &  3425   & 57565   & 12686    & 59618       \\
$m_{A} $         &  3402   & 57189   & 12603    &  59229     \\
$m_{H^{\pm}}$    &  3426   & 57565   &  12686    &  59619     \\

\hline
$m_{\tilde{\chi}^0_{1,2}}$
                 & 791, 800  &  989, 994 & 870, 873 & 1061, 1063   \\
$m_{\tilde{\chi}^0_{3,4}}$
                 & 1009, 1893 & 1424, 2705 & 1962, 3633 & 3070, 5703 \\

$m_{\tilde{\chi}^{\pm}_{1,2}}$
                 & 820, 1857  &  1038, 2619 & 901, 3570 & 1113, 5557\\
$m_{\tilde{g}}$  & 4964    &  7399 & 9009 & 14286\\

\hline $m_{ \tilde{u}_{L,R}}$
                 & 8894,8839  & 56887, 57050 & 15042, 14898 & 59898, 59990 \\
$m_{\tilde{t}_{1,2}}$
                 & 5874,6820  & 33309, 46976 & 10067, 12811 &35365, 49562\\
\hline $m_{ \tilde{d}_{L,R}}$
                 & 8895,8833 &  56887, 57077 & 15042,14898 & 59898, 60007\\
$m_{\tilde{b}_{1,2}}$
                 & 6780, 7340& 46856, 56860 & 12763, 14506 & 49431, 59701\\
\hline
$m_{\tilde{\nu}_{1}}$ 
                 &  8087     & 56842 & 13429 & 59259\\
$m_{\tilde{\nu}_{3}}$
                 & 6919      &  56842 & 13192 & 59085\\
\hline
$m_{ \tilde{e}_{L,R}}$
                &  8085, 8000   & 56815, 56796 & 13426,13247 & 59231, 59141 \\
$m_{\tilde{\tau}_{1,2}}$
                &  5374, 6911  &  56564,56688 & 12764, 13187 & 58796, 59048\\
\hline

$\sigma_{SI}({\rm cm}^2)$
                & 1.71$\times 10^{-45}$ & 6.64$\times 10^{-46}$ & 1.3$\times 10^{-46}$
				& 4.36$\times 10^{-47}$   \\

$\Omega_{CDM}h^2$
                & 0.114      & 0.123 & 0.115 & 0.125\\
\hline
\hline
\end{tabular}
}
\caption{ Four benchmark points for the red and blue points, which are consistent with
the dark matter relic density bounds. Sparticle and Higgs masses are in GeV units.
\label{table1}}
\end{table}
%


\section{Conclusion}

We have studied the supersymmetric GUTs with gravity mediated supersymmetry breaking in details. 
First, considering the dimension-six proton decay via heavy gauge boson exchange,
we pointed out that the supersymmetric GUTs with
 GUT scale $M_{GUT}$ up to $1.778\times 10^{16}$~GeV can be probed at the Hyper-Kamiokande experiment.
Second, for the supersymmetric GUTs with $M_{GUT} \ge 1.0\times 10^{16}$~GeV and 
$M_{GUT} \ge 1.2\times 10^{16}$~GeV, we showed that the upper bounds on the universal gaugino mass
are  $7.2$ TeV and 3.5 TeV, respectively, and thus the corresponding upper bounds on gluino mass
are  15 TeV and 8 TeV,  respectively.  In particular, the supersymmetric GUTs with
$M_{GUT} \leq 1.2\times 10^{16}$~GeV can be probed at the Hyper-Kamiokande experiment,
and   the supersymmetric GUTs with $M_{GUT}\ge  1.2\times 10^{16}$~GeV can be probed at 
the  ${\rm FCC}_{\rm hh}$ and SppC experiments via gluino searches. 
Thus, the  supersymmetric GUTs with gravity mediation can be probed by the ${\rm FCC}_{\rm hh}$, SppC,
and Hyper-Kamiokande experiments. In our previous study, we have shown that
the supersymmetric GUTs with anomaly and gauge mediated supersymmetry breakings are well within the reaches of 
these experiments. Therefore, we propose the concrete scientific goal for 
the ${\rm FCC}_{\rm hh}$, SppC, and Hyper-Kamiokande experiments: probing the supersymmetric GUTs.


\begin{acknowledgments}

This work is supported in part 
by the National Key Research and Development Program of China Grant No. 2020YFC2201504, 
 by the Projects No. 11875062 and No. 11947302 supported by the National Natural Science Foundation of China,
by the Key Research Program  of the Chinese Academy of Sciences, Grant NO. XDPB15,
as well as by the Scientific Instrument Developing Project of the Chinese Academy of Sciences, Grant No. YJKYYQ20190049.

\end{acknowledgments}



\end{document}